\documentclass[amsmath,amssymb,twocolumn,nofootinbib, superscriptaddress]{revtex4-1}

\usepackage[utf8]{inputenc}
\usepackage{times,bbm}
\usepackage{amsfonts}
\usepackage{xcolor}
\usepackage{mathtools}

\usepackage{graphicx}
\usepackage{float}
\usepackage{csquotes}
\usepackage{comment}

\usepackage{hyperref}
\hypersetup{
	breaklinks,
	colorlinks,
	linkcolor=gray,
	citecolor=gray,
	urlcolor=gray,
}

\usepackage{dsfont}

\newcommand{\ket}[1]{\left \vert #1 \right \rangle}

\newcommand{\braket}[2]{\langle #1 \vert #2 \rangle}

\definecolor{jens}{rgb}{.2,0.7,.9}
\definecolor{cadmiumgreen}{rgb}{0.0, 0.42, 0.24}

\definecolor{augustine}{rgb}{0,0,1}

\begin{document}

\title{Time evolution of
many-body localized systems in two spatial
dimensions}

\author{A.~Kshetrimayum}
\affiliation{Dahlem Center for Complex Quantum Systems, Freie Universit{\"a}t Berlin, 14195 Berlin, Germany}
\affiliation{\mbox{Helmholtz Center Berlin, 14109 Berlin, Germany}}

\author{M.~Goihl}

\affiliation{Dahlem Center for Complex Quantum Systems, Freie Universit{\"a}t Berlin, 14195 Berlin, Germany}

\author{J.~Eisert}

\affiliation{Dahlem Center for Complex Quantum Systems, Freie Universit{\"a}t Berlin, 14195 Berlin, Germany}
\affiliation{\mbox{Helmholtz Center Berlin, 14109 Berlin, Germany}}

\date{\today}

\begin{abstract}
Many-body localization is a striking mechanism that prevents interacting quantum systems from thermalizing. The absence of thermalization behaviour manifests itself, for example, in a remanence of local particle number configurations, a quantity that is robust over a parameter range. Local particle numbers are directly accessible in programmable quantum simulators, in systems of cold atoms even in two spatial dimensions. Yet, the classical simulation aimed at building trust in quantum
simulations is highly challenging. In this work, we present a comprehensive tensor network simulation of a many-body localized systems in two spatial dimensions using a variant of an iPEPS algorithm. The required translational invariance can be restored by implementing the disorder into an auxiliary spin system, providing an exact disorder average under dynamics. We can quantitatively assess signatures of many-body localization for the infinite system: Our methods are powerful enough to provide crude dynamical estimates for the transition between localized and ergodic phases. Interestingly, in this setting of finitely many disorder values, which we also compare with simulations involving non-interacting fermions and for which we discuss the emergent physics, localization emerges in the interacting regime, for which we provide an intuitive argument, while Anderson localization is absent. 
\end{abstract}

\maketitle

\section{Introduction}

While generic ergodic
systems are expected to thermalize under closed system evolution 
\cite{1408.5148,PolkovnikovReview,christian_review}, {constituting their own heat bath}, 
systems which exhibit \emph{many-body
localization (MBL)} are a robust exception to this 
paradigm \cite{Basko,RevModPhys.91.021001,PalHuse,christian_review}.
Such systems do equilibrate, but retain too much
memory of the initial condition so that the 
time averaged states could be described by a thermal
ensemble, due to \emph{localization}.
The localization gives rise to quasi-local constants of
motion in real space 
\cite{PhysRevX.5.031032,PalHuse,GoihlNew}, which need to be included in an equilibrium ensemble, leading to a non-{thermal equilibrium} state. MBL {can be seen} as an intricate 
generalization of the well-known
Anderson localization in which disorder and
interactions come together.
Since its 
discovery in the early years of this millennium \cite{Basko}, a plethora
of theoretical works followed {elucidating the 
rich and multi-faceted phenomenology of MBL in one spatial
dimension, ranging from a logarithmic growth of 
entanglement 
\cite{1304.4605,Pollmann_unbounded,Znidnaric2008,EisertOsborne06}
over slow information 
propagation \cite{PhysRevB.96.174201,1412.5605} to an 
\emph{area law} for the entanglement entropy
\cite{AreaReview}
for
highly excited eigenstates \cite{Bauer,1409.1252}. 
{Experimental realizations} have followed for
MBL systems in one spatial dimension
\cite{2DMBL, Smith2016,Dimitris,MBLScience}, 
corroborating some of the phenomenology.}

In two spatial dimensions, MBL is significantly less understood.
Experiments with ultra-cold atoms have been pursued \cite{2DMBL}, showing localization under
precisely controlled conditions. Yet, much of the phenomenology
is less clear -- to the extent that it has been suggested
that MBL may be unstable altogether
and that ergodicity could eventually be restored, albeit on very long time scales 
\cite{Roeck2017, RoeckImbrie}.
{Such assessments are made difficult by numerical treatments
being excessively challenging \cite{2DMBLBH}. Steps have been taken in the numerical analysis:
Ref.\ \cite{Wahl2017} constructs a two-dimensional cellular automaton,
further seminal work discuss finite  \cite{Kennes2018}  and
infinite \cite{Claudiusevolution} disordered systems numerically, while 
Ref.~\cite{Heyl2DMBL} targets weakly interacting systems of finite sizes. Exact diagonalization
limits discussions to} either non-interacting or extremely small systems.
{Tensor network approaches are immensely challenged by the 
entanglement build-up, even if this is slower compared to
ergodic systems  \cite{1304.4605,Pollmann_unbounded,Znidnaric2008}.
Still, given the unfavorable scalings of bond dimensions to faithfully
present quantum states as  tensor networks, this still gives rise to a
challenging and intricate state of affairs.}

In this work, we present a new take on the 
problem of simulating time
evolution of many-body localized 
two-dimensional quantum systems.
We discuss the physics of infinite two 
dimensional systems featuring discrete disorder 
using \emph{infinite projected entangled pair states
(iPEPS)}, building upon a methodology recently
introduced in Refs.\ \cite{Claudiusevolution,LCRGSirker},
in turn building upon Ref.\ \cite{CiracdisorderTI}.
The translational invariance inherent in this ansatz
is here be restored by exploiting a quantum dilation that embodies the classical disorder in giving rise to exact disorder averages; 
an ansatz suggested some time ago
\cite{CiracdisorderTI} and recently implemented for 
disordered two-dimensional systems \cite{Claudiusevolution}
in a proof-of-principle methodological study,
using a different iPEPS 
update from the one simple update employed here. 
While the so-called full update is known to be more accurate for ground state simulations for
the same bond dimension, whenever possible, it is an interesting observation in its own right that 
simple updates -- {more resource efficient procedures} 
-- turn out to be significantly more stable in time evolution algorithms,
as experience with numerical procedures has shown  \cite{Claudiusevolution} and is
convincingly confirmed in this work, presumably for being better able to reflect local changes in time evolution.
That is to say, our scheme that we employ here is more stable, resource
efficient and provides better control over the dynamics. For this reason, we have been able to achieve 
the longest available times in 2D dynamics {following a global quench} for strongly interacting systems ($t=3J$) {to date} in the thermodynamic limit, thanks to the disorder {present}.

We argue that while disorder averages are comparably feasible in one-dimensional
studies, it is such a two-dimensional setting for which quantum dilations to
capture classical disorder averages is particularly practical and relevant. Intriguingly, the implementation of programmable 
discrete disorder can avoid the issue of ergodic bubbles
right from the outset \cite{Roeck2017, RoeckImbrie}, sidelining the issue of stability of many-body localization in higher dimensions.
{Such an implementation of discrete disorder gives rise to a situation that is already intriguing 
in the non-interaction case reflecting Anderson localization.
Building upon early work
\cite{DiscreteAnderson}, there is a recent revitalized interest in rigorous studies
of Anderson localization for instances of discrete disorder in the absence of interactions 
\cite{DiscreteAnderson,ImbrieAndersonDiscrete,RandomSchroedinger,DiscreteAnderson,ImbrieAndersonDiscrete,DingSmart,LiZhang}. These rigorous results prove localization in specific regimes of discrete disorder
discussed in more detail later. Interestingly, 
within the settings considered here, however,
we do not find 
signatures of dynamical localization on the time scales considered.
For this phenomenon, we provide an explanation
in terms of discrete disorder leading to an effective hopping problem on every level. 
We augment this argument by numerical simulations of a finite
non-interacting system using exact diagonalization {which are further supplemented by iPEPS}}.
In the presence of interactions, we find signatures of
localization in the local particle number and 
{suitable Renyi entropies}, entering a highly exciting new physical regime in the first place, which we discuss in 
great detail.

{We will start by discussing the underlying paradigmatic model that is at the heart of our analysis, and then turn to discussing the numerical methods we make use of and develop to study} the disordered model (both the free fermions and iPEPS). We present the results for the non-interacting as well as the interacting instance of the Hamiltonian. In the method section, we will {specifically} describe how the {translationally} invariant iPEPS can be used to realize disorder by introducing {dilations}. The results section includes a discussion of the absence of Anderson localization and numerical evidence supporting it from two independent techniques. We then discuss the results for the evidence of many-body localization in the interacting case. Based on the particle imbalance $\mathcal{I}$ which we compute for different configurations of the parameters, we are able to estimate a crude dynamical phase diagram of MBL in 2D. The critical disorder strength is found to be $h\approx 6$ with at least 
four levels of disorder. We close by summarizing the results and giving an outlook for future work including possible experimental realizations in state-of-the-art analog quantum simulators.

\section{Model and localization measure}
The model we focus on 
is the spin-1/2 XXZ-Hamiltonian on a square lattice with disordered fields
\begin{equation}
    H = \sum_{\langle i,j \rangle} \left(S^x_iS^x_j + S^y_iS^y_j + \Delta S^z_iS^z_j \right)
    + \sum_i h_i S^z_i,
    \label{Hori}
\end{equation}
where $S^x$, $S^y$ and $S^z$ are the different Pauli spin operators {associated with a} particular site. $\Delta$ is the strength of the anisotropy, which we either choose to
be $\Delta=0,1$, which toggles many-body interactions. The value of the
magnetic field at a particular site is given by $h_i$. Usually $h_i$ are drawn
randomly from a continuous interval $[-h,h]$ for each site in the lattice,
but we will soon turn to other discrete probability measures.

The essence of MBL, so one can say, is the localization of its constituent particles leading to a breakdown of conductance \cite{Basko} and thermalization \cite{2DMBL} despite the presence of many-body interactions. A proxy for these effects is the local particle number dynamics {following} a quench from an particle imbalanced initial state. We consider a N{\'e}el state vector of the form
\begin{equation}
  |\psi_{0}\rangle_p = | \uparrow ,\downarrow ,\uparrow ,\downarrow ,\cdots ,\rangle\,.
\end{equation}
When subjected to the Hamiltonian evolution of a thermalizing Hamiltonian, the local particle imbalance quickly evens out and evolves towards a homogeneous particle distribution \cite{Trotzky}. However, if the Hamiltonian localizes the constituent particles, the initial particle imbalance will be measurable for very long times \cite{2DMBL}.
We stress that the observation of a remaining particle imbalance for a finite time window does not give information about the ``genuine'' quantum phase the system is actually in, as for long times the system can still thermalize
\cite{Roeck2017,abanin2017,Suntajs2019}. 
However, even localization for short times can be relevant for experimental realizations \cite{2DMBL} and practical 
applications such as quantum memories \cite{harnessing2019}. 

\section{Setting}
Usually, when working with disordered systems numerically, in order to obtain disorder averaged quantities simulations need to be run multiple times and the disorder average of the expectation values of the local observables are then calculated. In this case, a single realization of a system is not translation invariant and hence finite.
There is another technique of realizing disordered models that circumvents the above finite size effects and running the simulations multiple times to obtain statistics for the disorder average. The method makes use of additional auxiliary dilation 
spaces at every site whose spin states in superposition that upon tracing out this degree of freedom, one obtains the exact disorder averages, as introduced in Ref.\
\cite{CiracdisorderTI}. Since the combined system is translation invariant, we can access the thermodynamic limit using translationally invariant algorithms. This is the approach we will be taking in this work. We will describe them in more detail in the subsequent sections.

\subsection{Set up for iPEPS} \emph{Projected entangled pair states (PEPS)}
  are the generalization of \emph{matrix product states} 
  to higher dimensions \cite{PEPSOld,Verstraete-PRL-2006}. Similar to its one dimensional counterpart, PEPS target the physically relevant corner of the Hilbert space that is distinguished by its low entanglement content while representing a quantum state in higher dimensions \cite{Orus-AnnPhys-2014,VerstraeteBig,EisertTensorNetworks} that are of physical interest. One of the many advantages of such tensor network techniques is that they can directly study systems in the thermodynamic limit, thereby overcoming finite size effects, that one would often encounter using techniques like exact diagonalization. In this context, the 
  \emph{infinite projected entangled pair states (iPEPS)} \cite{iPEPSOld} have become the state of the art numerical tool in simulating two dimensional systems. They have been known to provide excellent variational ground state energies, in some instances even outperforming the state-of-the-art \emph{quantum Monte Carlo} calculations \cite{CorboztJ}. The success of iPEPS lies beyond simulating two dimensional simple cubic lattices. It has found applications in finding ground states of frustrated systems \cite{Xiangkhaf,ThibautspinS,KshetrimayumkagoXXZ} and realistic materials \cite{SSlandpeps1,SSlandpeps2,Kshetrimayummaterial}. They have also been used to describe thermal states in 2D \cite{piotr2012,piotr2015,Kshetrimayumthermal,Mishratwo-bodyBH} as well as steady states of dissipative systems \cite{Kshetrimayum,Weimereview}. While most of these works target the fixed points of the model, it is also possible, in principle, to use iPEPS for studying dynamics of a system. This is limited to only short time scales due to the fast growth of entanglement. The situation is true for all Tensor networks and even more severe for two dimensional systems further limiting the accessible time scales \cite{piotrevolution,Claudiusevolution}. In this work, we will use iPEPS to study the dynamics of the XXZ-Hamiltonian (Eq.\,\eqref{Hori}) in the presence of disorder and look for signatures of localization in different regimes of the anisotropy $\Delta$, as well as number of discrete levels $d_A$ of disorder. 
  
  In our setting, we exploit what can be called ``quantum parallelism'' to realize disorder in our translationally invariant system in the thermodynamic limit 
  as first proposed in Ref.\,\cite{CiracdisorderTI} 
  and realized in one \cite{LCRGSirker,LCRGSirker2018}
  and two \cite{Claudiusevolution} dimensional
  disordered systems.
  In essence, the method implements discrete disorder using auxiliary spin-$S$ systems for otherwise translational invariant Hamiltonians. 
  There is one of these auxiliary   spaces for each real space site and they are prepared in a superposition state of all their spin states. By adding another term
  to the Hamiltonian that projects these values onto the real space, we obtain discrete disorder landscapes. When calculating expectation values of observables, the states of the auxiliary space actually conveniently implement the disorder average over all possible disorder realizations.
  We will now break down this procedure into three important steps in order to implement this type of disorder.
  
  \begin{itemize}
  
  \item {\it Initialization:}
  We initialize our physical state
  {vector} $|\psi_{0}\rangle_p$ as a product that is easy to prepare experimentally, more specifically the N{\'e}el state, i.e.,
  \begin{equation}
  |\psi_{0}\rangle_p = | \uparrow {,} 
  \downarrow {,}\uparrow {,}\downarrow {,}\cdots \rangle    .
  \end{equation}
  For our simulations, we have chosen
  an iPEPS with a two-site unit cell and a
  checkerboard pattern as shown in Fig.\ \ref{ipepsinit}(a). This is sufficient to realize the configurations of interest. We also initialize the {auxiliary} state in a product state of equal superposition state {vector} $|+\rangle$, i.e.,
  \begin{equation}
  |\psi_{0}\rangle_a = |+{,}  + {,}+ {,}\cdots \rangle 
  . 
  \end{equation}
  For a spin-$S$ system, this superposition is given by {$|+\rangle = (2S+1)^{-1/2}
  (\sum_{s}|s\rangle)$}, where $s$ are the allowed spin states. Hence, the number of discrete values, our disordered field takes is $2S+1$ where $S$ is the spin of the {auxiliary}   space. Thus, the \emph{number of discrete levels of disorder}, which we 
  \emph{refer to as} $d_A$ is 2 for a spin-1/2, 3 for for a spin-1 {auxiliary}   system and so on. We then take the tensor product of the initial physical state {vector} and the initial {auxiliary} state {vector} and define this to be our overall initial state from where we 
  start quenching, i.e.,
  \begin{equation}
      |\Psi_0 \rangle = |\psi_{0}\rangle_p \otimes |\psi_{0}\rangle_a
  \end{equation}
  {where}
  $|\Psi_0 \rangle$ is a product state {vector}
  and hence an iPEPS with bond dimension $D=1$. This completes the initialization protocol, which we also illustrate in Fig.\,\ref{ipepsinit}.
  
  \begin{figure}
	\includegraphics[width=.47\textwidth]{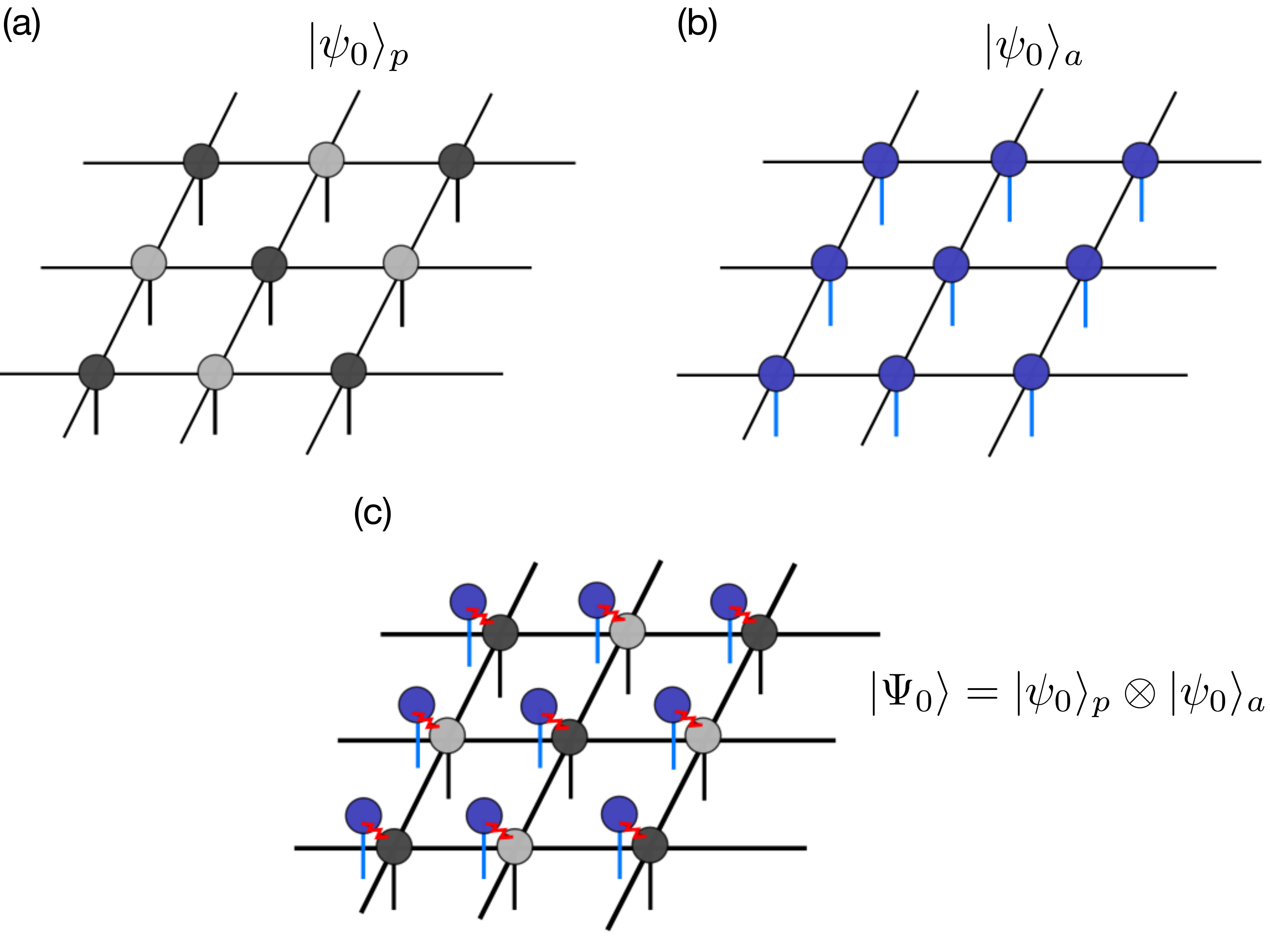}
	\caption{Initial state expressed in terms of iPEPS for (a) the physical state {vector} $|\psi_0 \rangle_p$ which is a N{\'e}el state, (b) the {auxiliary}   
	state {vector} $|\psi_0 \rangle_a$ is a product of equal superposition states and (c) the overall initial state
	{vector} $|\Psi_0 \rangle$, the tensor product of the previous two states. The red patterns correspond to the classical interaction between the physical and the 
    {auxiliary} states which is required for introducing the disorder. All the three states are iPEPS with bond dimension $D=1$ and the lattice extends indefinitely in all the directions. A choice of a two-site unit cell in a checkerboard pattern is enough to exactly represent this configuration.
	}
	\label{ipepsinit}
\end{figure}
  
  \item {\it Quench:}
  Once our initial state has been prepared, we perform the real time evolution of our disordered Hamiltonian. For this, the original Hamiltonian in Eq.\,\eqref{Hori} needs to rewritten 
  as 
  \begin{equation}
    H = \sum_{\langle i,j \rangle}\left( S^x_{i_p}S^x_{j_p} + S^y_{i_p}S^y_{j_p} + \Delta S^z_{i_p}S^z_{j_p} \right)
    + h\sum_i S^z_{i_p}S^z_{i_a}
    \label{Hanc} 
 \end{equation}
where the first term of the Hamiltonian is the sum over all the nearest-neighbor physical sites.
The second term couples each physical spins with its auxiliary spin but there is no coupling between different sites. This term projects the disorder contained in the auxiliary space onto the physical state using the local $S^z_{ip}S^z_{ia}$ coupling. $S^z_{ia}$ is defined such that the values of the disordered fields are taken from a fixed interval $[h,-h]$ with uniform distribution. Thus, for $d_A$, the values will be $h$ and $-h$, for $d_A=3$, it would be $h$, $0$ and $-h$ and so on. We will study the effect of disorder as we increase the dimension of the auxiliary system $d_A$ thereby allowing more levels of disorder configurations. We use the simple update \cite{simpleupdatejiang} to do a real time evolution of our modified Hamiltonian starting from the initial state vector $|\Psi_0 \rangle$,
\begin{equation}
    |\Psi(t) \rangle = e^{-iHt} |\Psi_0 \rangle .
\end{equation}
This update scheme is not only efficient, but also more stable while dealing with such non-equilibrium problems \cite{Claudiusevolution}. This might be due to the fact unlike the full update technique, the simple update does not require to compute the ill-conditioned norm tensor at every step.

\item {\it Readout:} 
Once we {have generated} 
the {state vector}
$|\Psi(t) \rangle$ 
using the procedure described above, we can compute the {expectation values} of suitable 
local observables. Such expectation values are already the exactly disorder averaged expectation value of all the possible configurations by construction. 
This can be easily seen from the following calculation
\begin{equation}
\begin{split}
    \mathbb{E} \langle \hat{O}(t) \rangle  
    &= \langle \Psi(t)|\hat{O}|\Psi(t) \rangle\\
    &= \langle \Psi_0|e^{iHt}\hat{O}e^{-iHt}|\Psi_0 \rangle\\
    &= ( {}_a\langle \psi_0| \otimes {}_p\langle \psi_0|)e^{iHt}\hat{O}e^{-iHt}(|\psi_0 \rangle_p \otimes |\psi_0\rangle_a).
    \end{split}
\end{equation}
The on-site expectation value is calculated at the physical site as the auxiliary sites are traced out. We use an instance of a CTMRG algorithm \cite{ctmroman2012,ctmroman2009} for this purpose. We also use the same effective environment to compute the different Renyi entropies of the reduced density matrices.
\end{itemize}
Thus, the above procedure circumvents the need for having finite systems to realize disordered systems, at the same time avoiding the need for multiple simulations for different disorder configuration and taking their average.

\subsection{Set up for non-interacting fermions} 
In addition to the iPEPS simulations described above, we 
have also run some free
fermionic calculations reminiscent of
the non-interacting case $\Delta=0$ in a finite
system (note that the mapping is not exact due to the presence of Jordan-Wigner strings
in two spatial dimensions). 
Because the dynamics {is} only governed by the single particle sector, systems of
size $40\times40$ are {perfectly} accessible. Moreover, we can implement continuous disorder
for these simulations. In accordance with our iPEPS simulation, we {again} consider
a N{\'e}el initial state and evolve it in time. We measure the particle number on even and odd sites as a measure of localization \cite{2DMBL} as described above.
Here, we are in principle not 
restricted to any final time but we since we are interested in comparing the
results to the iPEPS simulations, we evolve up to a few tunnelling times 
by integrating Schr{\"o}dinger's equation.
Additionally, we can access the single particle eigenstates and single 
particle eigenenergies of these systems via exact diagonalization, which we
employ to calculate the inverse participation ratio, another measure of localization.

\section{Results}
\emph{Results for non-interacting $\Delta=0$ case.} In this section, we {present} results for the non-interacting case $\Delta=0$. 
In this regime, it is possible to solve larger two-dimensional systems exactly in the single particle space. It 
{has been} rigorously established that one and two dimensional 
systems localize for continuous disorder \cite{AndersonA,AndersonB,AndersonC}.
For discrete disorder the situation is more subtle. In fact, 
seminal work has solved the long-standing puzzle whether 
localization occurs in the first place in one spatial dimension
to the affirmative \cite{DiscreteAnderson}: Interestingly, 
for one spatial dimension, any probability measure that has support on more than a single point will lead to the Hamiltonian having pure point spectrum and exponentially decaying eigenfunctions
and hence localization, even though bounds to 
localization lengths are implicit. These results are
compatible with rigorous insights into dynamical localization
for suitable random 
Schr{\"o}dinger operators \cite{RandomSchroedinger}.
In higher dimensions, slightly 
weaker statements are shown, basically for sufficiently
large disorder \cite{DiscreteAnderson}, for 
disorder with sufficiently large numbers of discrete levels of disorder
\cite{ImbrieAndersonDiscrete}, or for parts of the spectrum
\cite{DingSmart,LiZhang}. These results apply equally well to our
situation of non-interacting fermionic systems.

\begin{figure*}
  \includegraphics[width=1\textwidth]{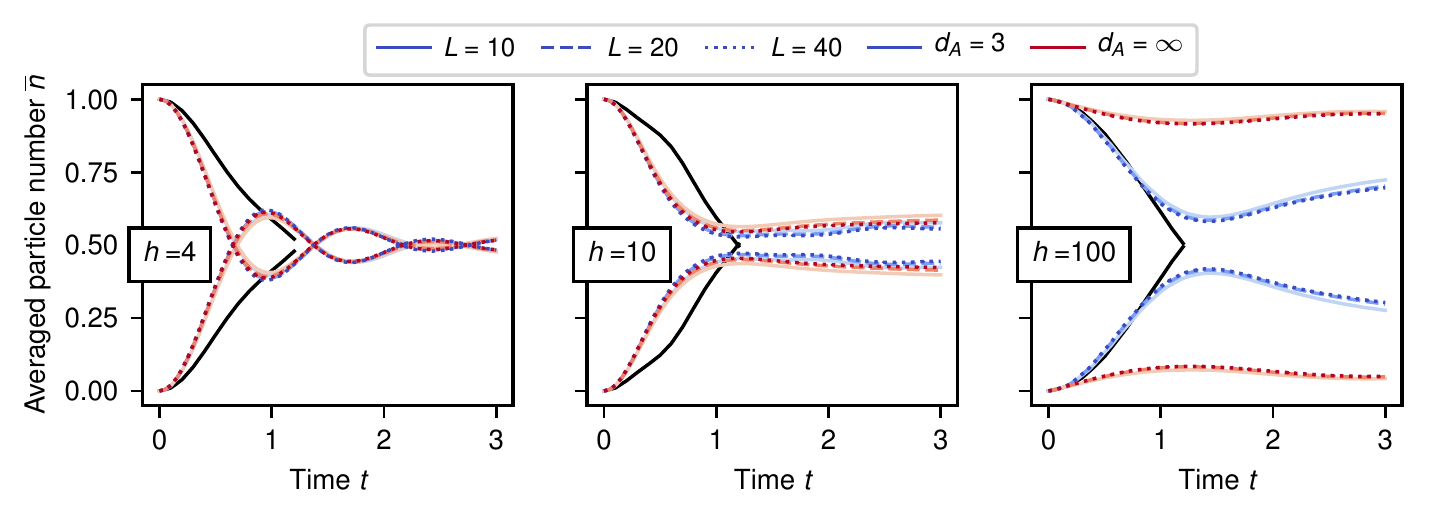}
  \caption{
    Averaged local particle number for even and odd sites in a
    free fermionic model for system sizes $L=10,20,40$ (markers). Averages are taken over $100$ realizations.
    Blue shades are for discrete disorder (spin-1) and red shades
    for continuous disorder. The black dotted lines are iPEPS results for infinite system. They are presented only to give a qualitative prediction of how the results using the free fermion simulation will change in the thermodynamic limit. A one-to-one quantitative comparison should not be made with the iPEPS results, since the plot shown is for $D=4$. However, the agreement between the two techniques is striking for the large disorder case $h=100$ since the entanglement growth is much slower here and the bond dimension do not play much role within the time scale presented here.}
  \label{f:free}
\end{figure*}

The dynamics of the particle number for even and for odd sites 
in the {non-interacting}
fermionic case is 
shown in Fig.\ \ref{f:free}. Here, we {present} results
for three disorder strengths ($h=4,10,100$) and two kinds of disorder: Continuous disorder
is shown in red shades and a three-level discrete (spin-$1$) 
disorder in blue shades. 
The two curves plotted depict the particle number for odd and even sites, respectively. Furthermore, we plot data obtained for the infinite system with the 
iPEPS code in black. This serves a more qualitative purpose however, since the plots shown are for iPEPS with fixed bond dimension $D=4$ and therefore we should be careful in making a one-to-one comparison with the exact 
diagonalization results quantitatively. 
For $h=4$, we find that the initial imbalance evens out on the time scales considered. 
There is no apparent difference for the two disorder models considered.
This apparent lack of localization
is by no means incompatible 
with the above
proven localization: On the one hand, 
in two spatial dimensions (unlike in one spatial dimension),
the disorder has to be sufficiently strong to encounter
localization. More importantly, on the other hand,
the figure of merit applied will
only encounter localization on the spatial extent of single lattice sites.
Hence, the absence of localization for the magnetization is
compatible with localization for longer localization lengths. In
fact, the machinery developed here gives rise to a tool
to explore this rich physics for discrete disorder in higher
spatial dimensions.

For $h=10$, we find a first 
{signature} of localization for the time scales considered as a weak imbalance -- signified by a gap between the two curves -- remains. 
When comparing the two disorder models{,} 
we already see a hint towards an observation
that will become more clear in the strongly disordered case. The continuous disorder 
results in a slightly larger gap. When we set $h=100$, there is a large gap for the
continuous disorder model{,} but only a small one for the discrete disorder model. The 
simulation for the infinite system agrees very well with the finite calculations
for $t<1$. It furthermore suggests {that} with increasing system size the gap closes completely.
Moreover, we find that increasing the levels of the discrete model results in a
larger gap (data not shown). 

\begin{figure}
  \includegraphics[width=0.45\textwidth]{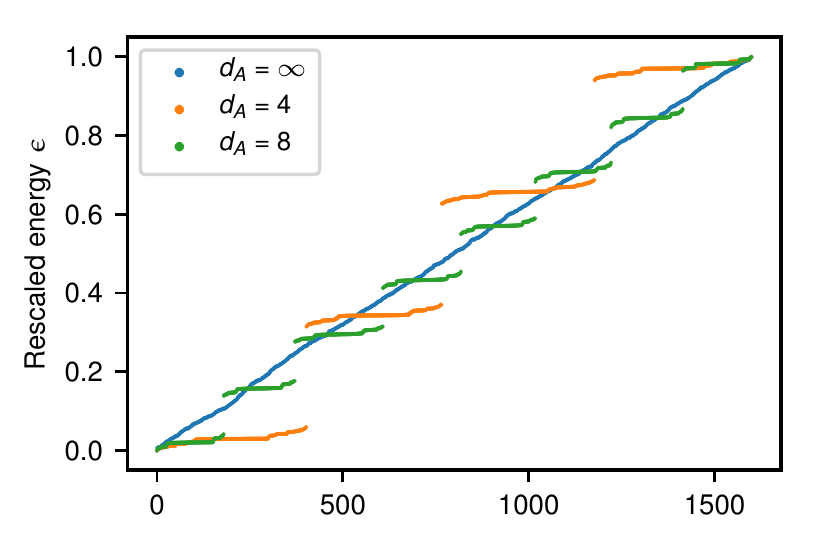}
  \caption{
    Single particle spectrum for the Anderson model with discrete and
    continuous disorder for $h=100$ averaged over 100 realizations.
  }
  \label{f:spec}
\end{figure}
\begin{figure}
  \includegraphics[width=0.45\textwidth]{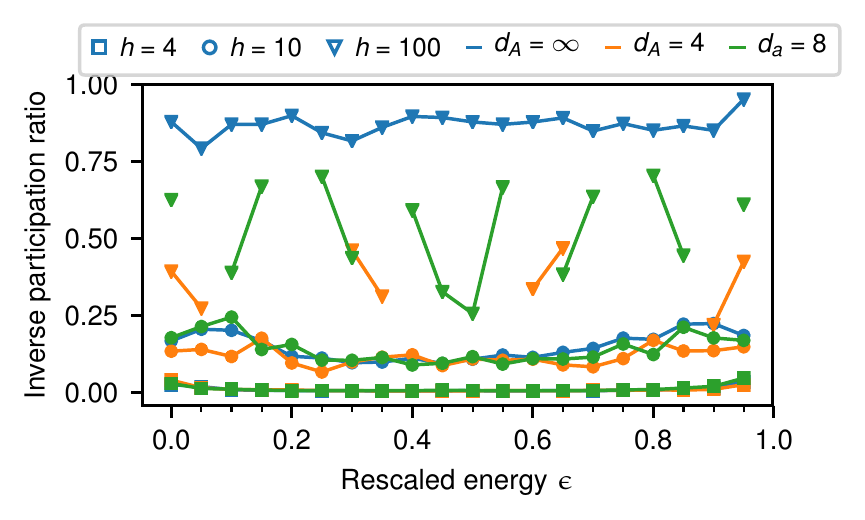}
  \caption{
    Cumulative inverse participation ratio for the Anderson model with discrete and
    continuous disorder for different disorder strengths. Lines are guides to the eye. When the energy levels for a certain value $\epsilon$ are not populated, no line is drawn.
  }
  \label{f:ipr}
\end{figure}
{To complement this analysis}, we also look at the single particle energy spectrum to understand the influence of the discrete disorder and why {dynamical} 
localization may not occur {for the observed times} 
in the discrete disorder model. In Fig.\,\ref{f:spec}, we plot the spectra for both models at high disorder $h=100$. We find that the spectrum for the continuous disorder is apparently still continuous. 
When discrete disorder with $s$ many levels is used, the spectrum is decoupled in
$s$ blocks, that have a weak bending caused by the hopping terms. 
This is compatible with the following intuitive explanation,
{which is furthermore in line with the above rigorous 
findings:}
Since the energy gaps between the levels are very large, the system effectively largely
decouples into sites of the same disorder strength. Depending on the position of the
next site with the same disorder value, the hopping strength will change, but
essentially the physics boils down to a hopping problem with a high coordination
number and
random hopping strengths. This implies that for long times, the system will evolve
towards a homogeneous state. 

To give more substance to this heuristics, 
we consider
the inverse participation ratio (IPR) defined as
\begin{equation}
    I_{\ket{E_k}} = \sum_i |\braket{i}{E_k}|^4\,,
\end{equation}
where $\ket{i}$ is a lattice site vector and $\ket{E_k}$ is the eigenvector with
corresponding energy $E_k$. This {provides} 
an estimate of the localization of the
eigenvectors in the following sense. If $\ket{E_k}$ only has support on a single
lattice site, its IPR is {unity}. If, {in contrast}, 
$\ket{E_k}$ has support on all
lattice sites, the IPR will be $1/L^2$. We consider a cumulative IPR for energy
segments. This means, we re-scale the spectrum according to
\begin{equation}
    \epsilon(E) = \frac{E-E_\text{min}}{E_\text{max}-E_\text{min}}\,,
\end{equation}
such that $0\leq\epsilon\leq1$. We then sum the IPR for all states in re-scaled energy intervals of size 0.05. The results are displayed in Fig.\,\ref{f:ipr}.
For low disorder $h=4$ (squares), the IPR is approximately the same for all three
types of disorder. For $h=10$ (circles), we see that at the ends of the spectrum,
the IPR is lower for less levels of disorder.
When considering the case of high disorder $h=100$, there is a strong qualitative
difference for the models. The continuous disorder results in a very high IPR 
throughout the full spectrum. Not only is the spectrum divided into blocks for the
discrete disorder, the resulting IPRs are also much smaller than in the continuous
case, indicating that these states are not localized. When including many-body 
interactions, these can be
interpreted as additional onsite fields that depend on the particle configuration.
This renders the potential experienced by the particles close to continuous 
restoring localization. We will explore this in the 
following section.

\subsection{Results for interacting $\Delta=1$ case} 
\begin{figure*}
	\includegraphics[width=0.9\textwidth]{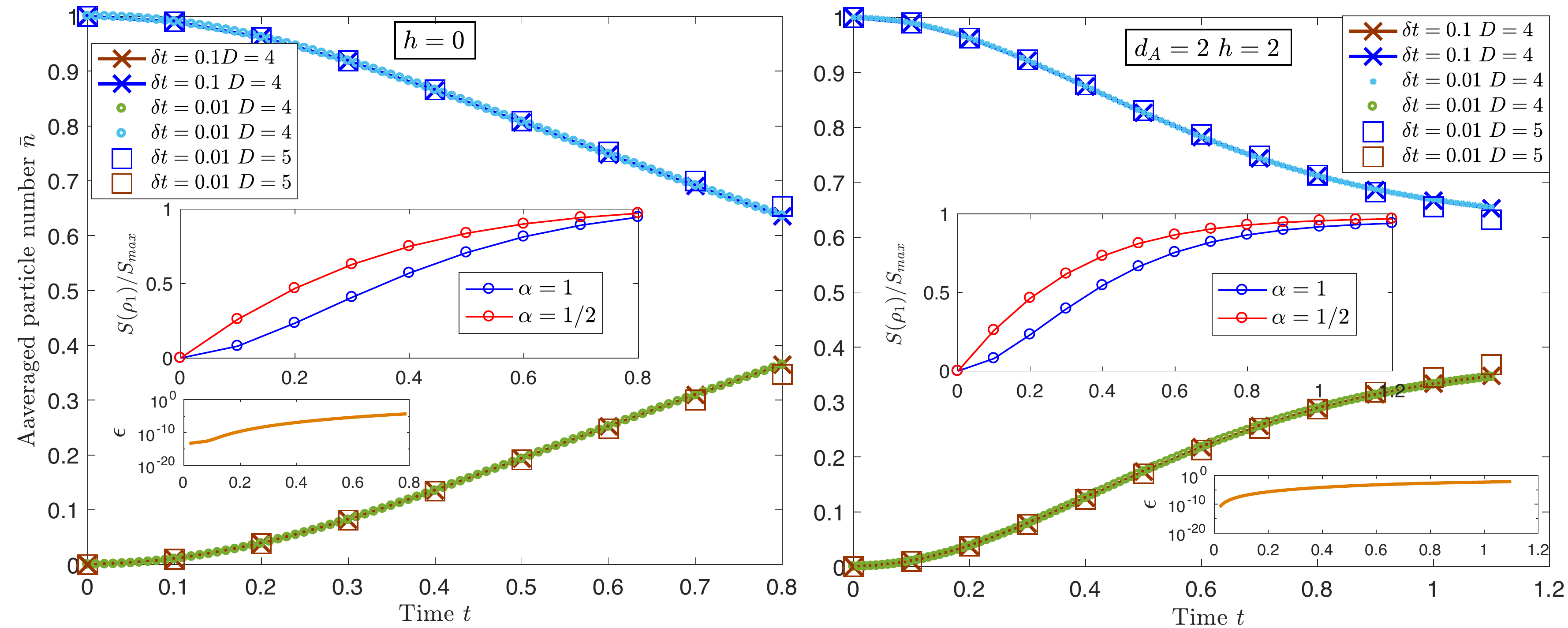}
	\caption{
		Real time evolution of the Heisenberg Hamiltonian starting from a N{\'e}el state. (Left) Expectation value of the particle number as a function of time for the two different sites for the case with no disorder $h=0$. [Big inset] Renyi entropies 
		of the reduced density matrix of one site as a function of time for $\alpha=1$ and $\alpha=1/2$. {The entropies start saturating to their maximum 
		value. [Small inset] Accumulated local truncation error of one site.}  (Right) The same evolution as above, but now with a disorder strength $h=2$ and for $d_A=2$. The simulation time has been extended slightly compared to the clean case without disorder. {One also notices a slow down in the growth of local Renyi entropies.} The simulations are done for $\delta t=0.1$ and $0.01$ for $D=4$ and $\delta t=0.01$ for $D=5$. The results are consistent with different Trotter sizes as well as different bond dimensions building trust into our simulations.}
	\label{ip2}
\end{figure*}

First, we will present the results for the clean case as well as the simplest case of disorder we can incorporate in our iPEPS simulations using the auxiliary   method. The simplest case is with binary disorder when the auxiliary 
system has a local Hilbert space ($d_A$) of two implying that our disorder landscape has two levels locally. We start by computing the expectation value of the particle number as a function of real time. The expectation values are computed at the two different physical sites of the tensor network. Since the initial physical state is a N{\'e}el state, 
its expectation values are one for the occupied site and zero at the empty site at $t=0$. As we initiate 
the quench, we want to analyze how the particle 
number changes with time. This is closely related to the 
experimentally used 
\emph{imbalance} {\cite{2DMBL,Trotzky,NumericalQuench2}}
which measures the difference of particle occupation between even and odd sites. In the absence of any disorder, this imbalance will eventually drop to zero or in other words, the particles will spread leading to a homogeneous particle distribution. This is shown in the left panel of Fig.\ \ref{ip2} although the time scale has been cut off early to avoid errors, according to criteria specified below.

\begin{figure}
	\includegraphics[width=0.48\textwidth]{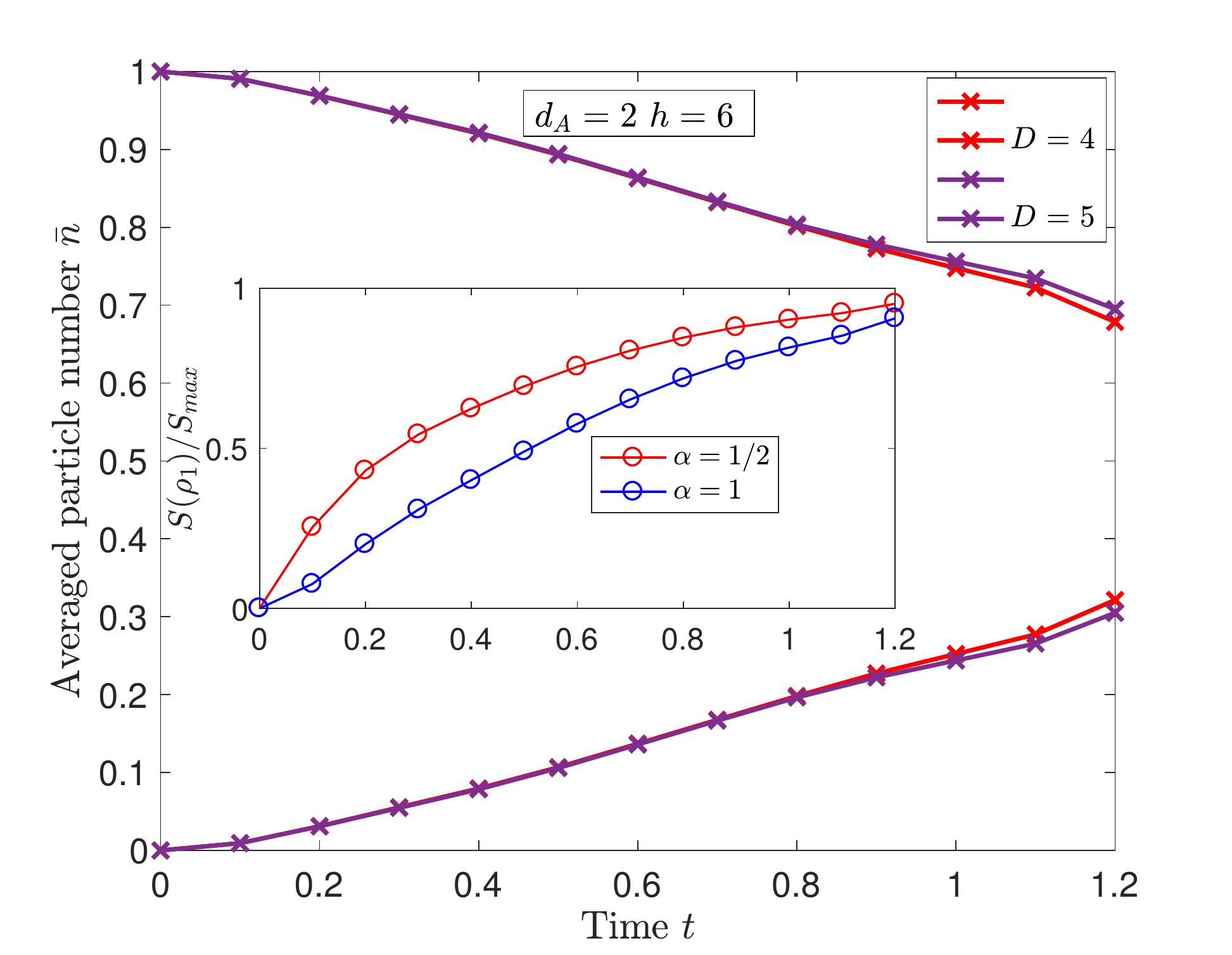}
	\includegraphics[width=0.48\textwidth]{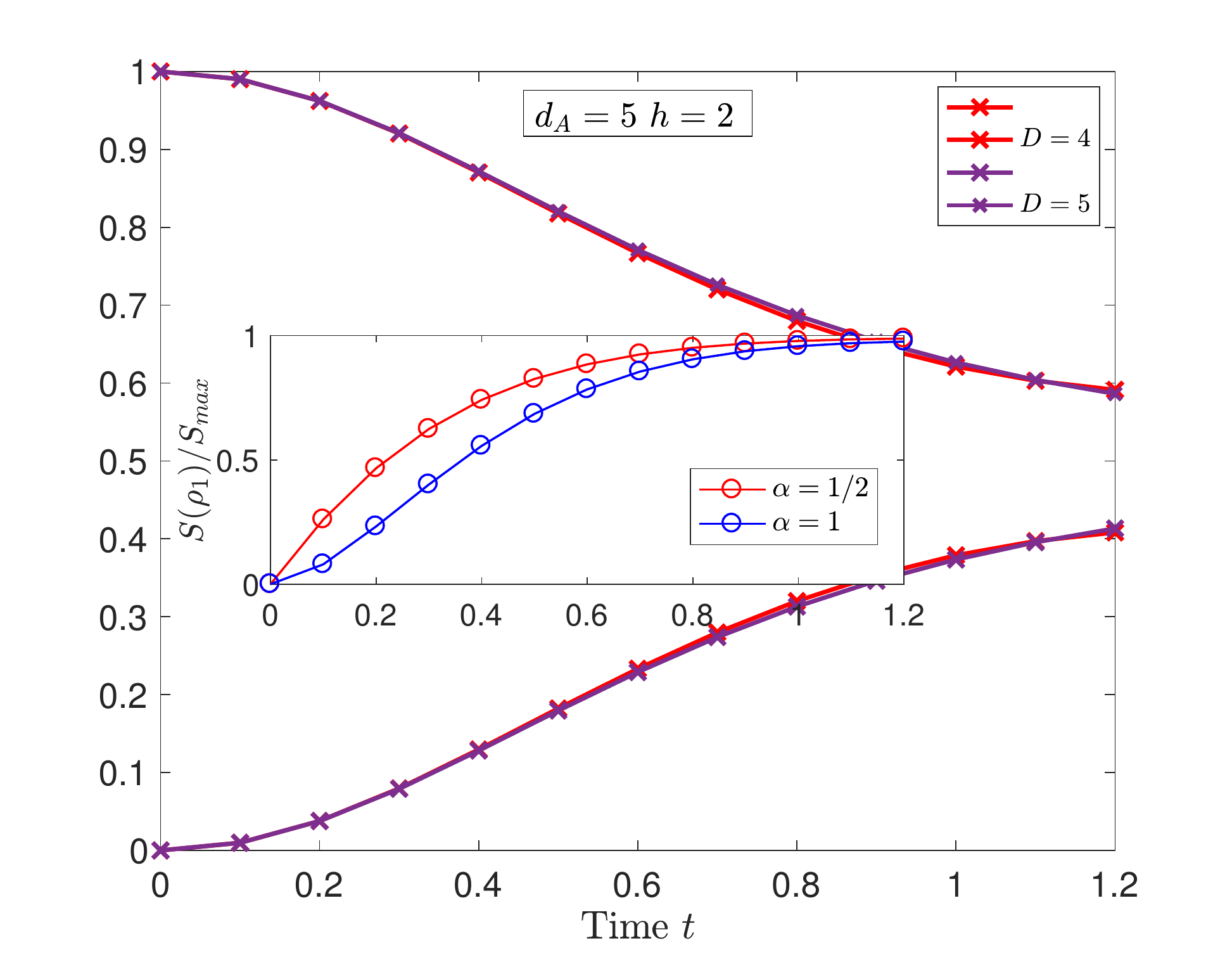}
	\includegraphics[width=0.48\textwidth]{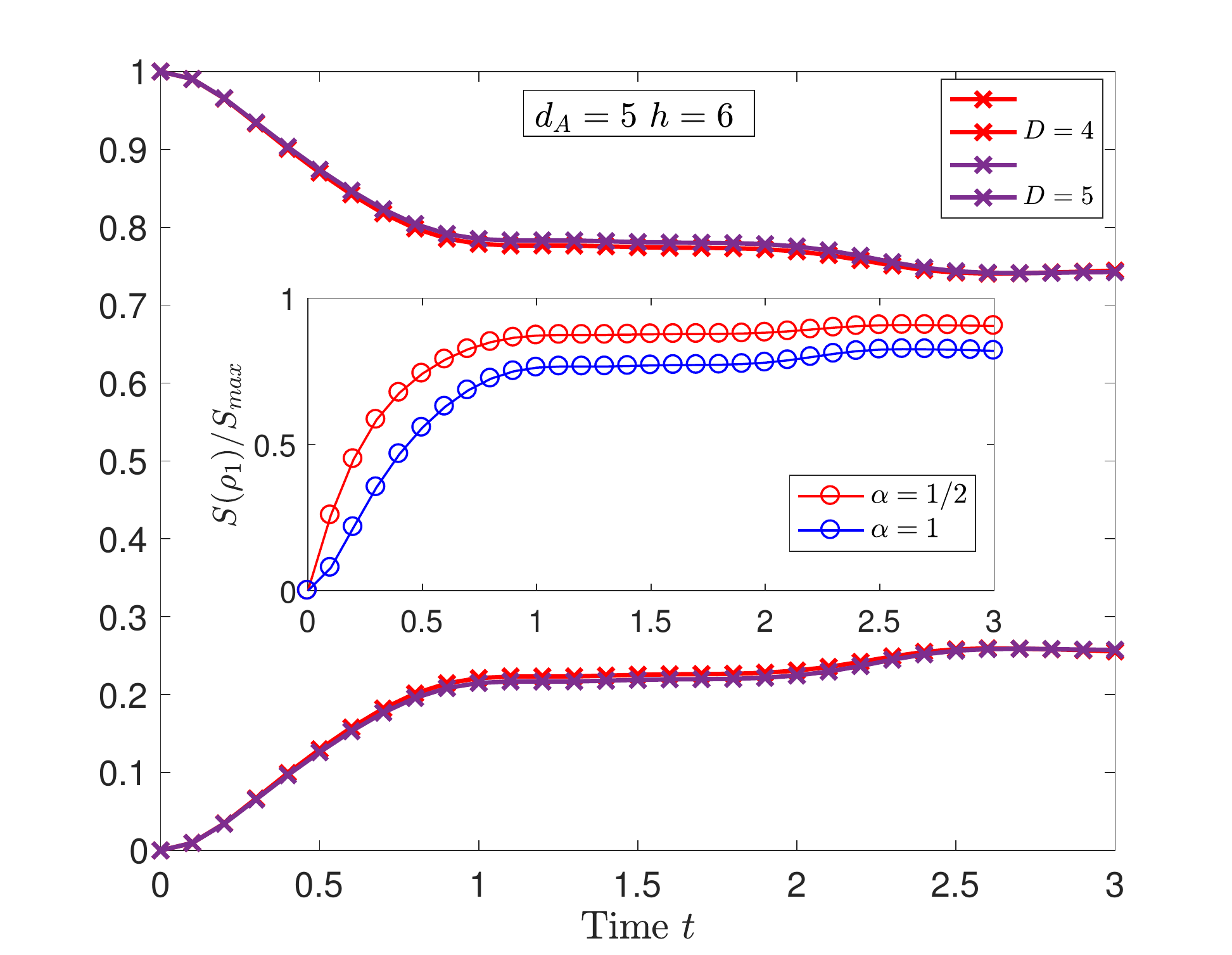}
	\caption{Similar to Fig. \ref{ip2} but for stronger disorder and more disorder levels. (Top) Relatively strong disorder $h=6$ but only two levels of disorder $d_A=2$. (Middle) More levels of disorder $d_A=5$ but weak disorder strength $h=2$. (Bottom) many disorder levels $d_A=5$ as well as relatively strong disorder $h=6$. In all the plots, we show simulations with bond dimensions $D=4$ and $5$ up to good agreement. This is also consistent with the growth of Renyi entropies which are shown in the insets.}
	\label{ahigher}
\end{figure}

For the calculations, we have used an iPEPS with a fixed bond dimension $D=4,5$ and Trotter step of $0.1$ and $0.01$. The reason for the comparably small bond dimension is that the physical dimension needs to be comparably large.
The results for both the Trotter steps as well as different bond dimensions are depicted in Fig.\ \ref{ip2} and Fig. \ref{ahigher}. As with all other tensor network approaches, iPEPS cannot be used for long time simulations due to the rapid 
\emph{growth of entanglement} \cite{SchuchQuench,AnalyticalQuench}, 
which can only be accounted for by ({in time exponentially}) 
large bond dimensions. This is a fundamental
challenge that can ultimately
not be overcome for any universal
classical simulation method, as 
Hamiltonian evolution is in principle as powerful
as a quantum computer (is {\tt BQP} complete in technical 
terms \cite{Vollbrecht}), and hence a universal classical efficient 
method of local Hamiltonian evolution 
for all times is unlikely to exist
\cite{DynamicalStructureFactors}. Using large bond dimensions in two dimensions is {significantly more challenging}
compared to the one dimensional case {and comes along
with significant computational effort}. {As a consequence,} 
the error measures {must necessarily be} 
less stringent here {compared to the situation in one spatial
dimension}.

The main criterion that we make use of for stopping the time 
evolution is a disagreement of the two largest available 
PEPS bond dimensions (which would here be $D=4$ and $D=5$),
reflecting a convergence in bond dimension: This convergence
builds trust in the expectation that higher bond dimensions
provide compatible results. This is shown in 
Fig.\ \ref{ip2} and Fig. \ref{ahigher}.
Some further intuition is also provided by monitoring the growth of the
local Renyi entropies in time starting from the initial product state.
{We compute the Renyi entropies $S(\rho_1)$ of order $\alpha=1$ and $\alpha=1/2$ for the reduced density matrix of one site. 
The scaling of Renyi entropies can be precisely
related to tensor network state approximations in one spatial dimension
\cite{Schuch_MPS}. Here, the issue at hand is more subtle, as we
operate in two spatial dimensions and
observing entropies of arbitrarily large
reduced states is excessively challenging. Still, the deviation from a saturated value
$S_\text{{max}}=\log(d_p)$, where $d_p$ 
is the 
dimension of the local Hilbert space of the physical 
spins, can be seen as an indication that the tensor
network approximation is still meaningful.
We provide these numbers in the inset of
Figs.\ \ref{ip2} and \ref{ahigher}.}
Furthermore, we show that our time evolution is stable against different 
Trotter steps $\delta t=0.1$ and $\delta t=0.01$, which is a key insight
in favour of the update scheme used here,
as Ref.\ \cite{Claudiusevolution} convincingly discussed issues with stability with full updates.
Along the way, we have also monitored 
the local truncation error, to see that it is not significant for our purposes, again shown in insets of Fig.\ \ref{ip2}. 
The local truncation error $\epsilon$ is the sum of the squares of discarded weights during the evolution for one site.
For the case without disorder, we plot the results for up to $t=0.8$ although $S(\rho_1)$ attains its maximal value at $t=1$ hopping strength. The local truncation error is of the order of $10^{-4}$ until this time).

We now introduce disorder to our system. For a disorder strength of $h=2$, we can see that the growth of entropy for a single site reduced density matrix slows down already, thereby allowing us to do time evolution to longer times. This is shown in the big 
inset of the right panel of 
Fig.~\ref{ip2}. Just like the previous case, $S(\rho_1)$ in this case also, becomes saturated after a few more time steps. The truncation error up to this time scale is of the order of $10^{-3}$ (shown in the small inset). Based on the particle number, there is still no strong indication of localization with such a weak disorder strength $h=2$ and low levels of disorder $d_A=2$. Increasing the bond dimension of the iPEPS will improve the simulation by a few time steps, but this is numerically very demanding. Similarly to the non-interacting case, we will investigate the influence of increasing the size of the local Hilbert space of the auxiliary  system, thereby allowing more levels of disorder locally as well as the disorder strength. 

{We first increase the disorder strength for the binary disorder case 
reflected by $d_A=2$. This is shown in the top panel of Fig.\
\ref{ahigher}. There is no significant change compared to the case 
of $h=2$ and $d_A=2$.
We now increase the number of levels of disorder in our system by increasing the local dimension of the Hilbert space of the auxiliary spins. We investigate this for $d_A=3,4,5,6$ and for different values of the disorder strengths
$h=2,4,6$. $d_A \xrightarrow{} \infty$, corresponds to the case 
of continuous disorder. {In Fig.\ \ref{ahigher}, we only show the plots for $d_A=2$, $h=6$ (top), $d_A=5$, $h=2$ (middle) and $d_A=5$, $h=6$ (bottom)}.}
{What we see from Fig.\ \ref{ahigher} is that merely 
increasing the disorder strength $h$ or the number of disorder levels $d_A$ 
alone is not sufficient to see signatures of localization, and
ergodicity seems to be preserved, judged from dynamical data. 
Only in the case with relatively strong disorder $h=6$ and many levels of disorder $d_A=5$ available, clear signatures of localization are encountered.
As a consequence, we are able to go to much 
longer times in our simulation $t=3J$. We would like to note here to the best
of our knowledge, this is the longest time achieved in time evolution with 2D tensor networks in the thermodynamic limit, facilitated by features of 
localization. Signatures of localization are also reflected by 
the considerable slow down of the growth of local Renyi entropies (as being
shown in the inset of the plots).}

\begin{figure}
	\includegraphics[width=0.48\textwidth]{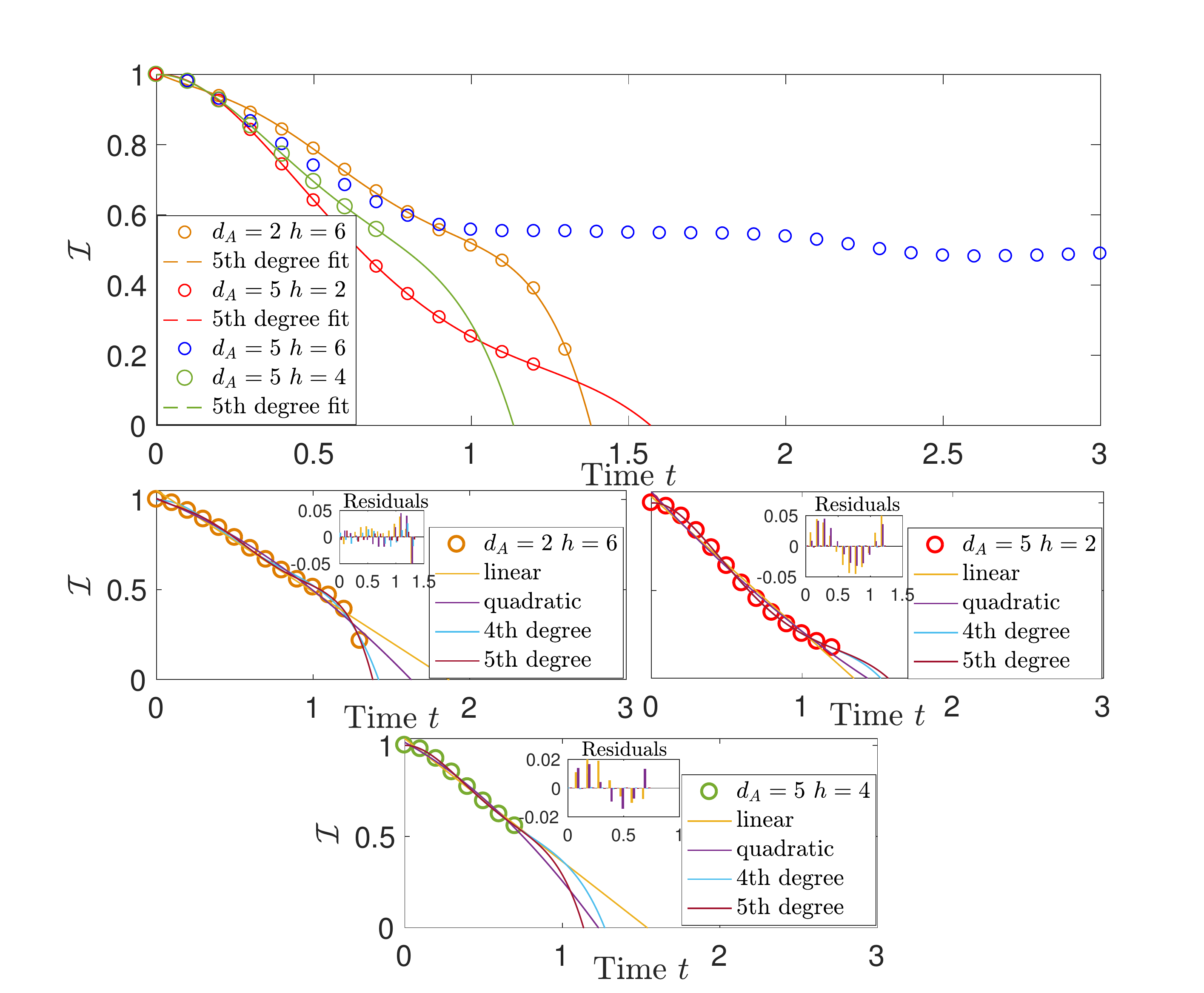}
	\caption{{Particle imbalance $\mathcal{I}$ for various configurations of disorder dimensions $d_A$ and strengths $h$. We show the dynamics of the longest available times for the localized case (blue circles, $h=6$ and $d_A=5$) which is {up to three hoppings}. Also shown are the cases where the particles do not localize (yellow circles with $h=6$, $d_A=2$, red circles with $h=2$, $d_A=5$ and green circles with $h=4$, $d_A=5$). The dynamics can be extrapolated using different polynomials such as linear, quadratic, 4-th and a 5-th degree fit and one can notice the imbalance dropping to zero in all these cases. Also shown are the residuals corresponding to each fit (dashed lines). The linear fit has the largest error while the 5-th degree polynomial fits in this sense most accurately.}}
	\label{imbal}
\end{figure}

{To be more comprehensive 
and systematic, we now consider different configurations of the disorder strength $h$ and disorder levels $d_A$ and plot the particle imbalance $\mathcal{I}$, defined as the difference in the occupation number of the two different sites. This is shown in 
Fig.\ \ref{imbal} for the configurations ($d_A=2$, $h=6$), ($d_A=5$, $h=2$), ($d_A=5$, $h=4$) and ($d_A=5$, $h=6$). As we see, only in the last configuration, one can go as far
as achieving the longest time evolution, because only then the system 
undergoes 
localization reflected by 
slow dynamics up to this time. For the
other situations, one has to be content with the available short time dynamics.
To make predictions with a reliably statistical basis, we have nonetheless
extrapolated these available times
using different polynomial fits such as linear, quadratic, 4-th and a 5-th degree polynomial 
least-square fits. This procedure
does allow for crude predictive
statements on future behaviour and indeed, the particle imbalance in all these cases conveniently and convincingly
drop to zero (reflecting no remaining imbalance). These are shown by dashed lines in Fig.\ \ref{imbal} along with the residuals of their fit to be precise.}

\begin{figure}
	\includegraphics[width=0.25\textwidth]{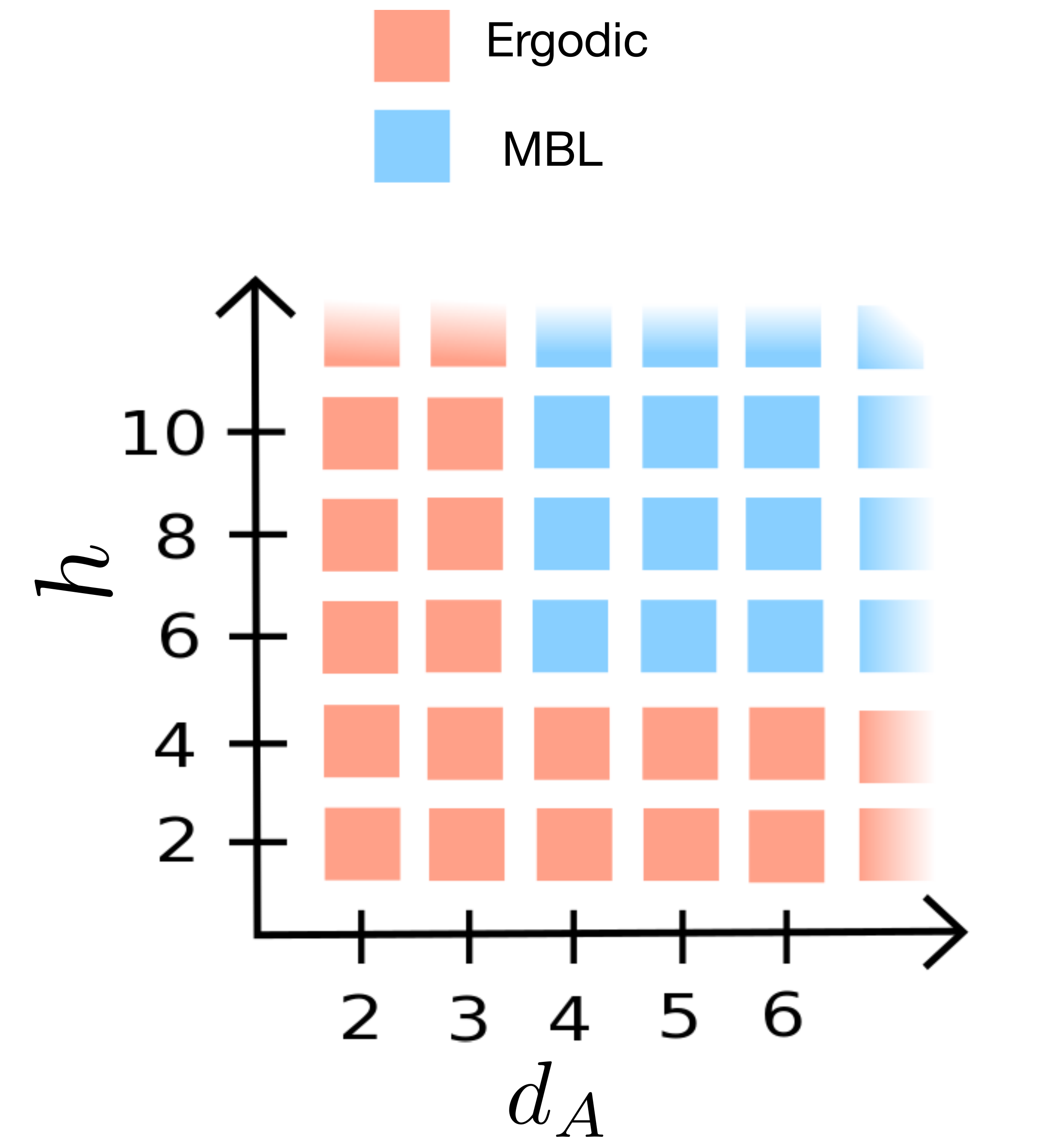}
	\caption{Crude estimate of the phase diagram as being assessed from
	dynamical localization as a function of the disorder dimension $d_A$ and the disorder strength $h$. The criterion to assign an ergodic or localized
	phase is whether the achievable simulation in time or the polynomial interpolation
	exhibits a localization of the imbalance in time or not.
	A disorder strength of $h=6J$ with at least four levels of disorder seems necessary to give rise to many-body localization in 2D.}
	\label{dynPD}
\end{figure}

{Based on the available information within the achievable times, we are 
now able to go a step further: Building on dynamical data, we can arrive at
crude estimates of the phase diagram of many-body localization in 2D based on the disorder strength $h$ and the levels of disorder $d_A$, judged from 
dynamical data. Even though these estimates are necessarily coarse-grained, it is still exciting
to see that the approach taken allows 
to draw conclusions along these lines,
in a regime that is very little 
studied using analytical and 
numerical state-of-the-art 
techniques. The results of this
endeavour are shown 
in Fig.\ \ref{dynPD}.}
{Pink boxes indicate that the system is likely to thermalize and is therefore ergodic, while blue boxes indicate the system localizes for the available time scales and is therefore in the MBL phase. This is the first dynamical phase diagram available for 2D dynamics with discrete disorder. Our dynamical phase indicates that in order to achieve MBL in two spatial dimensions, one needs a critical disorder strength of $h=6$ and disorder levels $d_A=4$. The experimental work of Ref.\
\cite{MBL2D} had found a critical disorder strength of $h=5.5$ for continuous disorder in an 
Aubrey-Andre model, even
though it is important to stress
that the underlying Hamiltonian model is that of a  two-dimensional
Bose Hubbard model. A
complementing theoretical 
work based on constructing 
cellular automata had found a 
critical disorder strength of $h=19$,
aimed again at exploring the 
disordered Bose-Hubbard model \cite{Wahl2017}}.

That is to say, we have been able to 
find that while discrete disorder landscapes
lead to no noticeable localization for the two dimensional
non-interacting systems we consider,
they appear to be capable of localizing interacting systems.
This is consistent with the argument given above that the interaction can be viewed as an additional source of randomness which depends on the adjacent particle configuration. It is also compatible with
the rigorous findings 
\cite{RandomSchroedinger,DiscreteAnderson,ImbrieAndersonDiscrete,DingSmart,LiZhang}
(as the disorder can be too 
small and the magnetization does not detect
a finite correlation length), {but adds scope to this, as we 
discuss the impact of specific small auxiliary dimensions}.
Before all, our findings can be seen as
an invitation to study in depth the rich physics of
discrete disorder beyond one spatial dimension.

\section{Conclusion and outlook} 
We have studied the effect of disorder in two dimensional system using two independent techniques: a free fermionic simulation for the non-interacting regime of the XXZ-Hamiltonian and an iPEPS algorithm for the interacting regime. By implementing discrete levels of disorder in the latter case as well as continuous disorder for the free fermionic case, we have found strong numerical evidence for the many-body localization in infinite two dimensional systems when using a sufficient number of disorder levels as well as disorder strength. Based on the dynamics of the particle imbalance for the available times, we have estimated a crude phase diagram of MBL in 2D, finding the critical disorder strength to $h=6$ and {at least} four levels of disorder $d_A=4$. Surprisingly, we do not find any evidence of localization for the infinite two-dimensional system for the non-interacting case using discrete levels of disorder, despite the mathematical proof of Anderson localization in two
spatial dimensions 
with continuous disorder. We have provided an intuitive argument on why this is the case based on a decoupling of potential levels which leads to an effective hopping problem, one that it at the
same time compatible with the findings of Ref.\
\cite{DiscreteAnderson}.
Our argument is supported by strong numerical 
evidence based on two independent techniques. 

We argue that the significance of our work is {four}-fold:
We present a stable numerical machinery that is
able to explore a regime of disordered lattice models in
higher dimensions that has formerly been significantly
less accessible. {Our machinery is more resource efficient, stable and provides better control over the dynamics. For this reason, we have been able to go to the longest available time scale of $t=3J$ in 2D, thanks to the disorder.} This is a technical, algorithmic improvement. 

Then, we are able to freshly explore the 
physics of discrete disorder \cite{DiscreteAnderson}, 
a regime that we think has received less attraction in the literature
than it deserves, giving the rich interplay of discreteness
of disorder and interactions, {and only
very recently is moving into the focus of attention in the
Anderson, i.e., non-interacting, case
\cite{RandomSchroedinger,ImbrieAndersonDiscrete,DingSmart,LiZhang}}. It would be very interesting to understand the interplay of discrete disorder also in view of stability of MBL and Griffiths effects.

{Excitingly, our tools are powerful enough to provide some
estimates of the phase diagram of many-body localization
assessed by investigating dynamical properties, even these
estimates are necessarily crude for time scales available. 
In light of the enormous difficulty of achieving such estimates, for example with quantum cellular automata \cite{Wahl2017},
we think that our dynamical method 
provides some handle on studying precise interplay and a 
phase diagram of the disorder strength and the number of 
levels of disorder with the system. 
The tools laid out here can be
seen as an invitation to quantitatively study this interesting regime more 
thoroughly. }

Finally, and maybe most importantly in the medium to long perspective,
we are able to provide benchmarks for \emph{quantum
simulators} \cite{BlochSimulation,CiracZollerSimulation} that are increasingly becoming available
in a number of physical platforms.
With the advent of programmable randomness,
this work can actually be probed directly in experiments as well.
As mentioned before, the programmable nature allows to 
avoid rare events of small local disorder and ergodic bubbles
leading to a potential instability \cite{Roeck2017, RoeckImbrie}
as a design principle for choosing disorder patterns.
{For example, the programmable,
re-configurable arrays of individually \emph{trapped cold atoms} with strong, coherent interactions realized by excitation to Rydberg states
\cite{ProbingQuantumSimulator} give rise to such a platform.
In systems of \emph{trapped ions} \cite{MonroeProgrammable}
and in \emph{superconducting devices} \cite{Dimitris},
large degrees of flexibility arise in programming 
potentials in one spatial dimension, settings in which
discrete disorder can be explored. Even beyond programmability,
the presence of one -- say, fermionic -- 
atomic species constituting \emph{discrete
disorder for another atomic species} 
\cite{PhysRevA.72.063616,PhysRevA.77.023601}
opens up interesting perspectives. 

Our work constitutes a basis on which a
compelling conclusion can be drawn
for the perspective of realizing such
programmable quantum simulators from
a complementing perspective: By
further developing and applying
 tensor network techniques,
we have entered a unprecedented regime for 
classical simulation techniques,
concerning the dimensionality of the system, the
way disorder is realized, and at the same time concerning the times reached. This 
information can be made use of to
build trust in the correctness of
an eventual programmable quantum
simulation in the sense of a 
partial \emph{certification}
\cite{BenchmarkingReview} of the quantum simulation. This will work for comparably
short times -- for long times, no
classical efficient computation will
be able to keep track of the quantum
dynamics \cite{Vollbrecht, DynamicalStructureFactors}. To access 
such long
times, one actually has to perform the quantum simulation in the laboratory, based on 
and guided by 
the insight the classical
simulation has provided. In this sense, our work can be viewed as a blueprint for a programmable quantum simulation using near-term quantum devices 
that accesses an
intricate quantum phase of matter.
It is the hope that the present
work stimulates such further endeavours.}

\section{Acknowledgements} 
We would like to thank N.\ Tarantino for pointing out the decoupling argument in the non-interacting case,
{A.\ H.\ Werner {and D.\ Toniolo} for helpful comments on the rigorous literature}, 
{and} D.\ M.\ Kennes and C.\ Hubig 
for discussions of their work. We would also like to thank 
M.\ Heyl for the suggestion to study the IPR for the 
{non-interacting} fermionic case. {This work has been supported by the ERC (TAQ), the DFG (CRC 183 project B01 and A03, FOR 2724, EI 519/7-1, EI 519/15-1), 
the Templeton  Foundation, and the FQXi.  This  work  has  also  received  funding  from  the  European  Unions  Horizon 2020  research  and innovation  programme  under  grant  agreement  
No.\  817482 (PASQuanS), 
specifically dedicated to programmable quantum simulators allowing for
programmable disorder.}


\section{Appendix} 

 In this work, we use \emph{exact diagonalization} (ED) and \emph{tensor network} methods. For the case of non-interacting system, we use ED up to system size $40 \times 40$. For the interacting system, we use \emph{infinite projected entangled paired states} combined with the quantum dilation technique discussed in the main text, directly in the thermodynamic limit. For optimizing the tensors, we use the \emph{simple update} scheme originally introduced for ground state calculations~\cite{simpleupdatejiang}. The reasoning for choosing this scheme over the \emph{full update} has been discussed in the main text already. For the update procedure, we use iPEPS with bond dimensions $D=4$ and $5$ with \emph{Trotter} steps $\delta t=0.1$ and $0.01$. The combined dimension of the physical and the auxiliary spins used in these simulations are $d=d_p \times d_A = 4$, $6$,$8$ and $10$. 

Once the tensors are optimized, we use the \emph{CTMRG} technique~\cite{ctmroman2009,ctmroman2012,ctmnishino1996,ctmnishino1997} to contract the full environment of the tensors, thus targeting the thermodynamic limit. 
The CTMRG algorithm computes the effective environment of a particular site by contracting the whole infinite 2D lattice except the site at which we want to compute the observables. For this, one needs to obtain a set of fixed point tensors that makes up this effective environment. Details on how we do this can be found in Refs.\,\cite{ctmroman2012,ctmroman2009,Kshetrimayumthesis}. The bond dimensions of the environment used are at least the square of the bond dimension of the ipeps ($\chi \geq D^2$) and are sufficiently well-converged. The agreement between the expectation values of the highest available bond dimensions is used as one of the criteria for stopping our time evolution.

\bibliographystyle{naturemag}

\end{document}